\documentstyle[preprint,aps]{revtex}
\begin{document}
\preprint{UM-P-95/59, RCHEP-95/16}
\draft
\title{MASSIVE ELECTRODYNAMICS AND THE MAGNETIC MONOPOLES}
\author{A.Yu.Ignatiev\cite{byline1} and G.C.Joshi\cite{byline2}}
\address{Research Centre for High Energy Physics, School of
Physics, University of Melbourne, Parkville, 3052, Victoria,
Australia}
\date{}
\maketitle
\begin{abstract}

We investigate in detail the problem of constructing magnetic monopole
solutions within the finite-range electrodynamics (i.e., electrodynamics
with non-zero photon mass, which is the simplest extension of the standard
theory; it is fully compatible with the experiment). We first analyze the
classical electrodynamics with the additional terms describing the photon
mass and the magnetic charge; then we look for a solution analogous to the
Dirac monopole solution. Next, we plug the found solution into the
Schr\"{o}dinger equation describing the interaction between the the
magnetic charge and the electron. After that, we try to derive the Dirac
quantization condition for our case.  Since gauge invariance is lost in
massive electrodynamics, we use the method of angular momentum algebra.
Under rather general assumptions we prove the theorem that the
construction of such an algebra is not possible and therefore the
quantization condition cannot be derived. This points to the conclusion
that the Dirac monopole and the finite photon mass cannot coexist within
one and the same theory. Some physical consequences of this conclusion are
considered. The case of t'Hooft-Polyakov monopole is touched upon briefly.

\end{abstract}

\section{Introduction}

Massive electrodynamics is electrodynamics in which the photon has a small
mass\footnote{Experimentally, the photon mass has to be very small: less than
$10^{-24}$ GeV or even $10^{-36}$ GeV, but the existence of such a bound is
not important for the purposes of this paper, so we do not go into detail
here.} rather than being exactly massless. It is perhaps the simplest and the
most straightforward extension of the standard quantum electrodynamics (QED)
\cite{GN,IZ}. It can be embedded into the standard $SU(2) \times U(1)$ model
\cite{clt}.

Although the introduction of a small photon mass may look unaesthetical, in
some respects massive QED is simpler theoretically than the standard theory.
For example, massive QED can be quantized in a manifestly Lorentz-covariant
way without introducing the indefinite metric while the standard QED cannot.
Also, the analysis of infrared properties of massive QED is easier because
there are no infrared singularities caused by the zero photon mass.

Note that although the photon is massive in the theory under consideration,
the electric charge is strictly conserved in massive QED as well as in
standard QED.  From the experimenatal point of view, massive QED is as
perfect as standard QED.

In massive QED the photon has three polarization states: two transverse ones
and one longitudinal. Despite this fact, it has been shown that the limiting
transition between both theories is in fact smooth rather than discontinuous
\cite{BS,N} . The physical reason is that the interaction of longitudinal
photons gets weaker as the photon mass tends to zero so that in the massless
limit they effectively decouple.

Such smoothness may even be elevated to a role of an important theoretical
principle: each physical phenomenon which can be described in standard QED
must have its counterpart description in massive QED; the two descriptions
must merge continuously in the limit of vanishing photon mass.

Independently of whether this conjecture is true or not, it is instructive to
see how it works (or fails) in various physical contexts since in this way a
better understanding of the physical situation can arise.

With this aspect in view, it is natural to consider those phenomena, for
which the gauge invariance of QED is crucial or at least  essential. One
example is the Dirac monopole \cite{D,rev}. Although there have been much
work devoted to various sides of magnetic monopole physics, the aspect we are
interested in has not been analyzed in detail so far.

The purpose of the present paper is to try and fill that gap. In other words,
we would like to know if magnetic monopoles and massive photons can coexist
within one and the same theory.

We know that the very existence of Dirac monopole is tightly connected with
the existence of gauge invariance of QED. In massive QED, there is no gauge
invariance anymore. So, the question is: what happens to magnetic monopoles
there? Do they survive the loss of gauge invariance?

Before starting any reasoning, we could just try to guess the right answer.
Obviously, it has to be either ``yes'' or ``no''. Let us consider these in
turn.

At first sight, the positive answer does not look impossible at all.  We know
that with the introduction  of photon mass the electrostatic field changes
from the usual Coulomb form, $E\sim\frac{1}{r^2}$, to the Yukawa form
$E\sim\frac{1}{r^2}e^{-mr}$. So, we would expect the magnetic field of a
monopole also to change from  $H\sim\frac{1}{r^2}$ to
$H\sim\frac{1}{r^2}e^{-mr}$. However, what would happen to the Dirac
quantization cony of Melbourne,
the magnitude of the photon mass and vanishing smoothly with the vanishing
photon mass? Or should the quantization condition remain the same in both
theories? Next, how can we make sure the string is unobservable if we do not
have gauge invariance?

 Now, if we guess that there can be no monopoles in massive QED, then again
there arise several questions. What about rotational invariance? We know that
the Dirac quantization condition can be obtained from rotational invariance
and angular momentum quantization \cite{6,7,8,9} without using gauge
invariance. Can we generalize that kind of arguments to the case of massive
QED? Next, what about the limit of vanishing photon mass? If magnetic
monopoles abruptly disappear when photon mass equals zero, then, how can we
make a continuous transition between the massive and massless QED?

So it seems that both  ``yes'' and ``no'' options offer interesting questions
to ponder on.

Our startegy in this paper will be as follows. First we write down the
classical Maxwell equations describing the magnetic charge and massive
photon. We then find the solutions which would go into the Dirac monopole
solution in the limit of vanishing photon mass (Section II). At this stage no
inconsistency arise , but no genuine monopoles either, because we have of
course the string attached to the monopole.

In the spirit of Dirac's approach we then turn to quantum mechanics in our
efforts to eliminate the mischievous string. We undertake this step in
Section III by plugging the solution found in Section II into the
Shr\"{o}dinger equation describing the interaction between the the magnetic
charge and the electron. After that, we try to derive the Dirac quantization
condition for our case.  Since gauge invariance is lost in massive
electrodynamics, we use the method of angular momentum algebra \cite{6},
generalized to the case of non-vanishing photon mass. Under rather general
assumptions we prove a theorem that the construction of such an algebra is
not possible and therefore the quantization condition cannot be derived. This
points to the conclusion that the Dirac monopole and the finite photon mass
cannot coexist within one and the same theory. Some physical consequences of
this conclusion, including the problem of continuity, are discussed at a
qualitative level in Section IV.    The case of t'Hooft-Polyakov monopole is
touched upon briefly at the end of the paper.

 \section{Classical Theory}

We start by writing down the Proca equation describing electrodynamics with
finite range (or equivalently with a non-zero photon mass)\footnote{We use
the Heaviside system of units throughout this paper and also put $\hbar =1,
c=1$.}:

\begin{mathletters}
\label{A}
\begin{eqnarray}
   & \partial^{\mu }F_{\mu \nu } = J_\mu  - m^2A_{\nu} & \label{A1}\\
   & \partial^\mu  \tilde{F}_{\mu \nu } = 0& \label{A2}\\
   & \partial^\mu  A_\mu  = 0, & \label{A3}
\end{eqnarray}
\end{mathletters}
where, the field strength $F_{\mu \nu }$ is connected with the 4-vector
potential
$A_\mu $ the usual way:

$$
         F_{\mu \nu } = \partial_\mu A_\nu  - \partial_\nu A_\mu .
$$

The dual pseudo--tensor $\tilde{F}_{\mu \nu }$ is defined as usual via

$$
       \tilde{F}_{\mu \nu } = \frac{1}{2}\varepsilon_{\mu \nu \alpha\beta}
F^{\alpha\beta};
$$
$J_\nu $ is the (electric) current density.  Next, the Maxwell equations
generalized to
include magnetic charge are:
\begin{mathletters}
\label{B}
\begin{eqnarray}
   & \partial^\mu F_{\mu \nu } = J_\nu &  \label{B1}  \\
   & \partial^\mu \tilde{F}_{\mu \nu } = J^g_\nu&  \label{B2}\\
   & F_{\mu \nu } = \partial_\mu A_\nu  - \partial_\nu A_\mu , \label{B3}
\end{eqnarray}
\end{mathletters}
where index $g$ denotes the magnetic current density.

Now the straightforward generalization of systems (\ref{A}) and (\ref{B})
reads \begin{mathletters} \label{C} \begin{eqnarray} & \partial^\mu F_{\mu
\nu } = J_\nu  - m^2A_\nu &  \label{C1} \\ & \partial^\mu \tilde{F}_{\mu \nu
} = J^g_\nu & \label{C2} \\ & \partial^\mu A_\mu  = 0 & \label{C3} \\ &
F_{\mu \nu } = \partial_\mu A_\nu  - \partial_\nu A_\mu. \label{C4}
\end{eqnarray} \end{mathletters} Let us rewrite this system in the
3--dimensional form: \begin{mathletters} \label{D} \begin{eqnarray} &
\nabla\cdot{\bf E} = \rho - m^2A_o     & \label{D1} \\ & \nabla\times{\bf E}
= - \frac{\partial{\bf  H}}{\partial t} - {\bf  j}_g& \label{D2} \\ &
\nabla\cdot{\bf  H} = \rho_g& \label{D3} \\ & \nabla\times{\bf  H} = {\bf  j}
- m^2{\bf  A} + \frac{\partial{\bf  E}} {\partial t}& \label{D4} \\ &
\frac{\partial A_o}{\partial t} + \nabla\cdot{\bf  A} = 0& \label{D5} \\ &
{\bf  E} = -gradA_o + \frac{\partial {\bf  A}}{\partial t}& \label{D6} \\ &
{\bf  H} = \nabla\times{\bf  A}.& \label{D7}
\end{eqnarray}
\end{mathletters}
This is the system of generalized Maxwell equations which would presumably
describe the existence of both magnetic charge and non-zero photon mass.

A few remarks are now in order.
\begin{enumerate}
\item The photon mass term
$m^2A_\mu $ in the right-hand side of Maxwell equations violates the symmetry
between the electric and magnetic charges.  This is clearly seen from the
comparison of Eq.~ (\ref{D1}) and (\ref{D3}): the "electric" equation
(\ref{D1} has as its solution the familiar Yukawa potential

$$
     {\bf E} = \frac{q}{4\pi r^3} e^{-mr}{\bf r},
$$
while the "magnetic" equation (\ref{D3}) does not feel the photon mass
at all.

\item The gauge invariance is completely lost due to the photon mass
term.  Indeed, it can be seen that the transformation
$$
    A_\mu  \rightarrow A_\mu  + \partial_\mu f
$$
is inconsistent with equations (\ref{A}) whatever the function $f$ is.
(Recall that the ordinary Maxwell equations in the Lorentz gauge,
$\partial_\mu A^\mu  = 0$, also do not allow gauge transformation with
the arbitrary function $f$.  However, these equations allow such
transformations for $f$ satisfying the condition $\Box f = 0$.  In
our case, even that restricted gauge invariance is lost.)  Note
that this loss of invariance has occurred already at the stage of the
Proca equations without magnetic charge and so it has nothing to do
with the introduction of magnetic charge.

\item Due to the loss of gauge invariance, the vector potential $A_\mu $
becomes observable quantity on the same footing as the field strength $F_{\mu
\nu }$.  It can be seen that the presence of photon mass term $m^2{\bf A}$ in
the righthand side of the equation of (\ref{D4}) creates a sort of additional
current density, in addition to the usual electric current ${\bf j}.$
\item
It is not immediately obvious that the loss of gauge invariance destroys the
consistency of the Dirac monopole theory and the validity of the quantization
condition. For example, the Aharonov-Bohm effect which is also based on the
electromagnetic gauge invariance, has been shown to survive in the massive
electrodynamics despite the absence of gauge invariance there \cite{BD}.
\end{enumerate}

Our modified Maxwell equations tell us that there arises the additional
magnetic field created by the "potential-current" $m^2{\bf A}$.  There
is no way to separate this additional magnetic field from the normal one.
Although in Proca theory (without magnetic charge) this circumstance
does not cause any problems, it becomes the main source of trouble once
magnetic charges are added to the massive electrodynamics, as we shall
see shortly.

After these general remarks, let us see if our system of "Maxwell + photon
mass +
magnetic charge" equations (\ref{D}) is consistent or not.  Let us try to
find a static monopole-like solution of that system.  For this purpose, we
assume the absence of electric fields, charges and currents (${\bf E}=0$,
$A_o = 0$, $\rho = 0$, ${\bf j}= 0$) as well as the absence of magnetic
current (${\bf j}_g=0$).

We then are left essentially with four equations

\begin{mathletters}
\label{E}
\begin{eqnarray}
   & \nabla\cdot{\bf H} = \rho_g&  \label{E1} \\
   & \nabla\times{\bf H} = -m^2{\bf A}& \label{E2} \\
    &{\bf H} = \nabla\times {\bf A}& \label{E3} \\
    &\nabla\cdot{\bf A} = 0. \label{E4}
\end{eqnarray}
\end{mathletters}
The first equation has the familiar Dirac monopole solution:

\begin{mathletters}
\label{F}
\begin{eqnarray}
   & {\bf H}^D = \frac{g}{4\pi r^3}{\bf r},& \label{F1} \\
   &A^D_r =\ A^D_\theta = 0,&  \label{F2} \\   & A^D_\varphi = \frac{g}
{4\pi r} \tan {\frac{\theta}{2}}.& \label{F3}
\end{eqnarray}
\end{mathletters}
As is well known, this solution involves a singularity in vector potential
along the line $\theta = \pi$ (``a string'').  Yet this singularity was
shown by Dirac to be only a nuisance without any physical significance.
\footnote{The magnetic charge can be assumed to be either scalar or
pseudoscalar under the action of P-parity (for more details see e.g. a
review \cite{ST}). In the former case, which we adopt in this paper, the
theory with magnetic monopoles is not parity invariant. However, none
of our
physical results would be changed if we treated the magnetic charge
as  a pseudoscalar (rather than scalar) quantity.}
Now if we plug this Dirac solution (or, more exactly, the Coulomb magnetic
field)
into the second equation, we immediately run into trouble, because clearly
$\nabla\times {\bf H}^D = 0$, instead of being equal to  $(-m^2{\bf A})$.
Let us try
to find a better solution by adding something to the Dirac solution.  In
this
way we write:
\begin{equation}
   {\bf H} = {\bf H}^D + {\bf H}',\; \; {\bf A}=  {\bf A}^D + {\bf A}',
\end{equation}
where the rotor and the divergence of the additional field ${\bf H}'$ must
satisfy
\begin{equation}
\label{eq:8}
      \nabla\cdot{\bf H}' = 0
\end{equation}
\begin{equation}
\label{eq:9}
      \nabla\times {\bf H}' = -m^2 ({\bf A}^D + {\bf A}'),
\end{equation}
while the divergence of the potential ${\bf A}'$ must vanish:
\begin{equation}
\label{eq:10}
      \nabla\cdot {\bf A}' = 0,
\end{equation}
because
$$
   \nabla\cdot{\bf   A}^D = 0  \mbox{ and }  \nabla\cdot ({\bf A}^D +
{\bf A}') = 0.
$$
Finally,
\begin{equation}
\label{eq:11}
      \nabla\times{\bf A}' = \nabla\times{\bf H}'.
\end{equation}

Now, we have the complete system of equations (\ref{eq:8}) through
(\ref{eq:11}) for
the rotors and divergences of both ${\bf H}'$ and ${\bf A}'$.

Let us see if there is any solution to it.  Taking the rotor of both
sides of Eqn.(\ref{eq:11}) and using Eqns. (\ref{eq:9}) and (\ref{eq:10})
we get the second-order
equation for ${\bf A}'$ only:
\begin{eqnarray}
\label{A}
   & (\triangle -m^2){\bf A}'  = m^2{\bf A}^D.&
\end{eqnarray}

The natural boundary condition for this equation is that ${\bf A}'$ must
vanish at infinity.  Note that after this equation is solved we have to
make sure that the transversality condition $\nabla\cdot {\bf A}' = 0$ is
obeyed.

In cartesian coordinates we get three decoupled scalar equations instead
of one vector equation:
\begin{equation}
\label{a}
      (\triangle - m^2)A_i' = m^2A^D_i \; i=x,y,z.
\end{equation}
The Green function for the equation
$$
    (\triangle - m^2)u = f,
$$
with the boundary condition of vanishing at infinity is:
$$
       G({\bf r}, {\bf r'} ) = \frac{1}{4\pi} \frac{exp (-m|{\bf r} -
{\bf r'} |)}{|{\bf r} - {\bf r'} |}.
$$
Therefore, we can write the solutions of the Eqns.(\ref{a}) as
$$
   A_x'({\bf r})=\frac{m^2}{4\pi}\int d^3 {\bf r'} A^D_x( {\bf r'} )
\frac{exp(-m|{\bf r} -  {\bf r'} |)}{|{\bf r} - {\bf r'}|},
$$
and the same for y, z components.  Thus we can write in the vector form:
\begin{equation}
\label{b}
{\bf  A'}({\bf r}) =\frac{m^2}{4\pi}\int d^3{\bf r'}{\bf A}^D({\bf r'})
\frac{exp\left(-m|{\bf r} - {\bf r'}|\right)}{|{\bf r} - {\bf r'}|}.
\end{equation}
Let us check that this solution is indeed transverse: find $\nabla\cdot
{\bf A}'({\bf r})$.

We have (here $ {\bf R} = {\bf r} - {\bf r'}$):
$$
   \nabla\cdot f(R){\bf g}({\bf r'}) = \nabla f(R) \cdot {\bf g}({\bf r'})
= f'(R) \cdot \nabla R \cdot {\bf g}({\bf r'}) = \frac{f'(R)}{R} {\bf R}
{\bf g}({\bf r'}).
$$
(Note that all differential operations are taken with respect to the vector
${\bf r}$.)

Next,
$$
f'(R)=\frac{d}{dR}\left(\frac{e^{-mR}}{R}\right)=-\frac{e^{-mR}}{R^2}(1 +
mR).
$$

Finally we obtain
$$
   \nabla\cdot {\bf A}'({\bf r}) = - \frac{m^2}{4\pi} \int d^3 {\bf r'}
\frac{e^{-mR}}{R^3}(1 + mR) {\bf R}{\bf A}_D({\bf r'}).
$$

Let us now show that it is zero.  Note that l.h.s (and r.h.s) are
{\em pseudoscalar}
(because both ${\bf A}'$ and ${\bf A}_D$ are {\em pseudovectors}).  To
construct this pseudoscalar, we have only two vectors at our disposal:
${\bf r}$
and ${\bf n}$ (the unit vector along the monopole string).
{}From them, we can make only two pseudoscalars: $({\bf r}
\times {\bf n}) {\bf n}$ and $({\bf r}\times{\bf n}){\bf r}$.
Both of then are zero (in addition to that, the first combination is ruled
out by the
condition that it must be linear in ${\bf n}$, but not quadratic).

Thus we have shown, that our solution satisfies both the equation
$(\triangle - m^2){\bf A}' = -m^2{\bf A}^D$ and the subsidary condition
of transversality, $\nabla\cdot {\bf A}' = 0$.

Let us now find the restrictions on the form of the potential ${\bf A}'
({\bf r})$ which follow from the general principles (dimensional analysis,
rotational
invariance and space reflection).

The most general form of ${\bf A}'({\bf r})$  , as a {\em pseudovector}
depending only on two vectors, ${\bf r}$ and ${\bf n}$, is:

\begin{equation}
  {\bf A}'({\bf r})    = \tilde{f} ( r, {\bf n} {\bf r} ) ( {\bf n} \times
{\bf r}).
\end{equation}
It is easier to work with dimensionless quantities, so let us write again
our initial formula
$$
 {\bf A}'({\bf r})  = \frac{m^2}{4 \pi } \int d^3 {\bf r'} {\bf A}^D
({\bf r'}) \frac{e^{-mR}}{R},
$$
and make a change of variables
\begin{eqnarray*}
   & {\bf r'}_1 = m{\bf r'}&\\
   & {\bf r}_1 = m{\bf r}&\\
   & {\bf R}_1  = m{\bf R},&
\end{eqnarray*}
so that ${\bf r'}_1$, $ {\bf r}_1$, ${\bf R}_1$ are all dimensionless.

Now,
\begin{eqnarray*}
   & d^3 {\bf r'} = \frac{1}{m^3} d^3 {\bf r'}_1&\\
   & {\bf A}^D ({\bf r'}) \sim \frac{1}{\bf r'} = \frac{m}{r'_1}&\\
   & R = \frac{R_1}{m}.&
\end{eqnarray*}
So that we get:
\begin{eqnarray*}
& {\bf A}' (\frac{{\bf r}_1}{m}) = \frac{m^2}{4 \pi }  \frac{1}{m} \int
d^3 {\bf r'}_1 \frac{exp(-R_1)}{R_1} \left(-\frac{g
}{4 \pi r'_1} \frac{{\bf n} \times {\bf r'}_1}{r'_1 - ( {\bf n}
{\bf r'}_1 )}\right)& \\
&=\ mgf( r_1, {\bf r}_1 {\bf n})
 ({\bf n} \times {\bf r}_1).&
\end{eqnarray*}
Now, go back to the old variables:
\begin{equation}
  {\bf  A} '({\bf r}) = m^2 gf(mr, m({\bf r}{\bf n})) ({\bf n} \times
{\bf r}).
\end{equation}
Now $f$ is a dimensionless function of two variables.

Let us check if this most general form satisfies the condition of
transversality, $\nabla\cdot {\bf A}' = 0$.
Use:
\begin{eqnarray*}
  & \nabla\cdot(S{\bf V}) = S \nabla\cdot {\bf V} + {\bf V} \nabla S &\\
  & \Rightarrow \nabla\cdot {\bf A'} = m^2 g \{ f \nabla\cdot ( {\bf n}
\times {\bf r})+  {\bf n} \times {\bf r} \nabla f\}.&
\end{eqnarray*}
But
$$
  \nabla\cdot ( {\bf V}_1 \times {\bf V}_2 ) = V_2 \nabla\times {\bf V}_1
-  {\bf V}_1 \nabla\times {\bf V}_2.
$$
Then,
$$ \nabla\cdot ({\bf n} \times {\bf r}) =0 . $$
Therefore
$$
   \nabla\cdot {\bf A}' = m^2 g  ( {\bf n} \times {\bf r}) \nabla f.
$$
Denote
$$
f(mr, m({\bf r} {\bf n})) = f(x, y)|_{x=mr, y= m({\bf r} {\bf n})}.
$$
  Then,
$$
   \nabla f = \frac{\partial f}{\partial x}  m  \cdot \nabla r +
\frac{\partial f}{\partial y} m \cdot \nabla ( {\bf r}{\bf n}).
$$
But
\begin{eqnarray*}
   \nabla r = \frac{{\bf r}}{r} , \,\, \nabla ({\bf r} {\bf n}) = {\bf n}.
\end{eqnarray*}
Therefore,
$$
  \nabla f = \frac{\partial f}{\partial x} \cdot m \cdot\frac{{\bf r}}{r}
+  \frac{\partial f}{\partial y}m \cdot {\bf n}.
$$
Hence we see that both terms $\sim {\bf r}$ and $ \sim {\bf n}$ give zero
when multiplied by the vector
product $( {\bf n} \times {\bf r})$.

\underline{Conclusion}: our general form for ${\bf A}'$ {\em always \/}
satisfies the transversality condition.

Now consider the problem of possible singularities in ${\bf A}'({\bf r})$.

In principle, there are two potential sources of singularities within
${\bf A}'({\bf r})$:
\begin{enumerate}
\item The $\frac{1}{R}$ behaviour of Green's function.
\item The singularity of string due to ${\bf A}^D$  factor.
\end{enumerate}

The worst case is when they occur simultaneously.  Let us consider this case.
That is, we consider the case,
when the observation point$(O)$ lies on the string, while the integration
point $(I)$ approaches it (i.e.
observation point).

To see what happens near the observation point, let us shift the integration
variables:
\begin{eqnarray*}
   {\bf r'} = {\bf r} - {\bf R} ,\,\, d^3 {\bf r'} = d^3 {\bf R}.
\end{eqnarray*}
Now choose the coordinates: see Fig.1.
Put the origin of the spherical system of coordinates at the point M (i.e.
where monopole is), z--axis
directed opposite  the string of monopole.  Call this system M.

Now, introduce a second spherical system with the origin at O.  Call this
system O.  Obviously, $d^3 {\bf R}$
has a simple form in the system O:
$$
  d^3{\bf R} = R^2 dR sin \theta_0 d\theta_0 d\phi,
$$
while the vector potential has a simple form in the system M:
$$
   {\bf A}_D (I)_{I\rightarrow 0} = \frac{g}{4 \pi r} \cdot tg
\frac{\theta_M}{2} \approx \frac{g}{4 \pi r } \frac{2}{\pi - \theta_M}
{\bf e}(\phi),
$$
where ${\bf e}(\phi)$ is a unit vector in the azimuthal direction.

Now, we need to find the relation between the angles $\theta_M$ and
$\theta_0$.

For this purpose consider $\triangle IOM$:
in it, we know two sides, $OM=r$ and $OI=R$, and one angle,
$\angle IOM=\theta_0$.  If we find the
angle, $\angle OMI = x$ then $\theta_M$ will be simply $\theta_M = \pi - x$.
We have:
$$
tg\,x = \frac{IH}{HM} = \frac{Rsin\,\theta_0 }{r - Rcos\,\theta_0} \approx
\frac{R}{r} sin\, \theta_0.
$$
Since
$x$ is assumed small, we obtain:
$$
   x \approx \frac{R}{r} sin\, \theta_0.
$$
But
$$
  {\bf A}'({\bf r}) \approx \frac{g}{4 \pi r} \frac{2}{x} {\bf e}(\phi)
= \frac{g}{2 \pi} \cdot \frac{1}{R sin\,\theta_0}\cdot {\bf e}(\phi).
$$onder on.
onder on.
onder on.
onder on.
onder on.

Putting all together, we get:
\begin{eqnarray*}
  & {\bf A}'({\bf r}) \approx \frac{g}{2 \pi } \int R^2 dR sin\,\theta_0
d\theta_0  d\phi \cdot \frac{1}{R} \cdot \frac{1}{R sin\,\theta_0} {\bf e}
(\phi)&\\
                     & \approx \frac{g}{2 \pi } \int dR  \int^{2\pi }_{0}
d\phi \, {\bf e} (\phi).&
\end{eqnarray*}
 From this form, it is clear that ${\bf A}'({\bf r})$ does not have any
singularity in the case considered.
 (Note also, that $\int^{2 \pi }_0 d\phi \,{\bf e} (\phi) = 0 $.)  But
because we have considered the {\em worst} possible
case, we may conclude that {\em there are no singularities in ${\bf A}'
({\bf r})$ at all.}

Let us now find the structural form of the magnetic field ${\bf H}'
({\bf r})$.

 From Eqn.(\ref{b}) which gives us the integral expression for the vector
potential ${\bf A}'$,
we can obtain the formula for ${\bf H}'$ by taking rotor:
\begin{eqnarray}
\label{bh}
   {\bf H}'({\bf r}) = \nabla\times {\bf A}' ({\bf r}) = \frac{m^2}{4 \pi}
\int d^3 {\bf r'} \frac{e^{-mR}}{R^3} ( 1 + mR) ( {\bf A}_D \times {\bf R} ).
\end{eqnarray}
Now, because ${\bf H}'({\bf r})$ is a vector (not a pseudovector!) depending
on the two vectors only, ${\bf n}$
and ${\bf r}$, it has the following most general form:
\begin{equation}
\label{eq:mandy}
   {\bf H}'({\bf r}) = h(r, {\bf n}{\bf r}){\bf r} + g(r, {\bf n}{\bf r})
{\bf n}.
\end{equation}
A question arises: can ${\bf H}'({\bf r})$ be spherically symmetric, that is,
can it take the
form
$$
    {\bf H}'({\bf r}) = h(r) \cdot {\bf r} \; \;?
$$

The answer is no, because in that case we would have
$$
   \nabla\times {\bf H}'({\bf r}) = 0
$$
everywhere, which is inconsistent with the initial Maxwell equations.

\section{Quantum Mechanics}

Having considered the classical theory of massive electrodynamics with
magnetic charge, we can now turn to quantum mechanics. Since 1931, the Dirac
quantization condition has been derived in many ways differing by their
initial assumptions. Obviously those methods using gauge invariance (such as
original Dirac derivation or the Wu-Yang formulation \cite{WY} ) are not
applicable in our case. Other methods \cite{6,7,8,9} depend on rotational
invariance and we can try to generalize them to include massive
electrodynamics, too. Before doing so, let us recall very briefly the essence
of the standard arguments.

Consider an electron placed in the field of a magnetic charge $g$. The
angular momentum operator for the electron is given by
\begin{equation} {\bf
L}={\bf r} \times (-i\nabla + e{\bf A}) + eg \frac{\bf r}{r} \; \; \; e>0.
\end{equation}
Despite the stange-looking second term, this operator can be
shown to obey all the standard requirements of a {\em bona fide \/}
angular
momentum: see commutation relations Eqs.~ (22--24) below. The angular
momentum is the generator of rotations. The two terms are individually
not angular momentum operators, in spite of the appearance of the first
term, but the sum is an angular momentum operator (see \cite{6}
 for more details). Moreover, $L_i$
commute with the hamiltonian
\begin{equation}
\label{eq:v}
 H = -\frac{1}{2m}{
(\nabla + ie {\bf A} )}^2 + V(r).
\end{equation}
Next, it was shown in \cite{6} that the quantity ${\bf L} \frac{\bf
r}{r}=eg$, should be quantized according to
\begin{equation} \label{dirac} eg= 0, \; \pm \frac{1}{2}, \; \pm 1
\ldots,
\end{equation}
which is the Dirac quantization condition.

Now, we would like to generalize this result to the case of massive
electrodynamics. Unfortunately, this turns out to be impossible: we will show
that the angular momentum operator cannot be defined for the system of charge
plus monopole within massive electrodynamics. More exactly, the following
theorem holds.

\underline{Theorem}

There are no such operators $L_i$ that the following standard properties are
satisfied:
\begin{equation}
\label{l1}
  \left[ L_i, L_j\right] = i \varepsilon_{ijk} L_k
\end{equation}
\begin{equation}
\label{l2}
  \left[ L_i, r_j \right] = i \varepsilon_{ijk} r_k
\end{equation}
\begin{equation}
\label{l3}
  \left[ L_i, D_j \right] = i \varepsilon_{ijk} D_k
\end{equation}
where

\begin{equation}
\label{l4}
  {\bf D} = -i\nabla +   e {\bf A}
\end{equation}
is the kinetic momentum operator,
\begin{equation}
\label{eq:vii}
   {\bf D} = m \dot{{\bf r}}
\end{equation}

Note that the conditions of this theorem are not too restrictive: for
example,
we do {\em not} require that the hamiltonian be rotationally invariant
(i.e.,
$ [ L_i, H]=0$ is not required).

\underline{Proof}

Our proof consists of two parts.  First we prove that from Eqns.~(\ref{l1}
--~\ref{l4})  it follows that
\begin{equation}
\label{L}
   \left[ L_i, H_j \right] = i \varepsilon_{ijk} H_k
\end{equation}
where ${\bf H}$ is the magnetic field,
$$
    {\bf H} = \nabla\times {\bf A}.
$$
Second, we show that Eqn.~(\ref{L}) is incompatible with the general form
for ${\bf H}$,
Eq.~(\ref{eq:mandy}).
Let us start with the first part.

Commuting Eq.~(\ref{l4}) with itself, we obtain
\begin{equation}
\label{l5}
  \left[ D_i ,D_j \right] = -i \varepsilon_{ijk} H_k
\end{equation}
Therefore, we have this expression for the magnetic field:
\begin{equation}
  H_k = \frac{i}{2} \varepsilon_{kij} \left[ D_i, D_j \right].
\end{equation}
We note that this form of the commutator implies non-associativity and the
appearance of the three-cocycle \cite{j}.

Now, let us plug this expression into the commutator $\left[ L_n, H_k
\right]$:
\begin{eqnarray*}
 & \left[ L_n ,H_k \right] = \left[ L_n, \frac{i}{2} \varepsilon_{ijk}
\left[ D_i, D_j \right] \right] = &\\
&\frac{1}{2} \varepsilon_{kij} (L_nD_iD_j - L_nD_jD_i - D_iD_jL_n +
D_jD_iL_n).
\end{eqnarray*}
We then commute $L_n$ and $D_j$ in the two middle terms of the above equation
using
Eq.(\ref{l3}):
\begin{equation}
\label{la}
    (...) = L_nD_iD_j - ( i \varepsilon_{njp} D_p + D_jL_n)D_i - D_i( -i
\varepsilon_{njp}D_p + L_nD_j) + D_jD_iL_n.
\end{equation}
The terms containing $\varepsilon$--tensor can be rewritten using
Eq.~(\ref{l5}):
\begin{equation}
\label{l6}
   i \varepsilon_{njp}(D_iD_p - D_pD_i) = -\varepsilon_{njp}
\varepsilon_{iqp}H_q = -\delta_{ni}H_j + \delta_{ji}H_n.
\end{equation}
Next, the rest of Eq.~(\ref{la}) can be transformed to contain only
$\left[ L D \right]$ and $\left[ D D \right]$
commutators.
\begin{eqnarray}
\label{l7}
   L_nD_iD_j - D_jL_nD_i - D_iL_nD_j - D_jD_iL_n &=& \left[ L_nD_i,
D_j\right] - \left[D_iL_n, D_j \right] \nonumber \\
                                                 &=& \left[ \left[ L_n,
D_i \right], D_i \right] \nonumber \\
                                                 &=& \left[ i
\varepsilon_{nir} D_r, D_j \right] \nonumber \\
                                                 &=& \varepsilon_{nir}
\varepsilon_{rjs}H_s  \nonumber \\
                                                 &=& \delta_{nj}H_i -
\delta_{ij} H_n.
\end{eqnarray}
Note that in obtaining Eq.~(\ref{l6})  and (\ref{l7}) we have used the
identity
\begin{equation}
   \varepsilon_{ijk} \varepsilon_{lmk} = \delta_{il} \delta_{jm} -
\delta_{im} \delta_{jl}.
\end{equation}
Finally, putting Eq.~(\ref{l6}) and (\ref{l7}) together, we finally obtain
Eq.~(\ref{L}).  Thus we have shown that indeed
Eq.~(\ref{L}) follows from Eqns.~(\ref{l1}--\ref{l4}).

Now we turn to the second part of our proof.  Let us insert the general
form of the magnetic
field ${\bf H'}$, Eq.~(\ref{eq:mandy}), into the commutator $ \left[ L_i,
H_l \right]$; we shall disregard the
${\bf H}^D$ field because it satisfies the correct commutation relations
with $L$ due to Eq.~(\ref{l2}):
\begin{eqnarray}
\label{eq:xiii}
     \left[ L_i, H_l' \right] =&&\ \left[ L_i, hr_l + gn_l \right]
\nonumber \\
                              =&&\ - i\varepsilon_{ijk}r_jn_kr_l
\frac{\partial h}{\partial ({\bf r}{\bf n})} + i\varepsilon_{ilp}r_p -
i\varepsilon_{ijk} r_jn_kn_l \frac{\partial g}{\partial ({\bf r}{\bf n})}.
\end{eqnarray}
If $L_i$ were good angular momentum operators, this had to be equal to
\begin{equation}
\label{eq:xiv}
   i\varepsilon_{ilq}H_q' = i \varepsilon_{ilq}(hr_q + gn_q).
\end{equation}
Now, compare Eqs.~(\ref{eq:xiii}) and (\ref{eq:xiv}).  We notice that:
\begin{enumerate}
\item terms proportional to $h$ in both equations coincide.
\item terms, quadratic in $n_i$ in Eq.~(\ref{eq:xiii}) must vanish
(because the remaining term in             Eq.~(\ref{eq:xiv}) is linear
in $n_i$ and the condition ${\bf n}^2 = 1$ has never been used),
therefore, we should have
$$
   \frac{\partial g}{\partial ({\bf r} {\bf n})} = 0.
$$
\item the remaining term (proportional to $n_k$) in Eq.~(\ref{eq:xiii})
should be antisymmetric with
respect to exchange $ i \leftrightarrow l$, in particular, it must vanish
for $i = l$. Taking
$i = l =1$; $j=2$, $k=3$, we see that this term reduces to
$$
   -2ixyn_z \frac{\partial h}{\partial ({\bf r}{\bf n})},
$$
whence
$$
  \frac{\partial h}{\partial ({\bf r}{\bf n})} = 0.
$$
It follows then that $g=0$, too.
\end{enumerate}

Thus, we obtain, that the angular momenta with correct commutation
relations can exist only if
${\bf H'}$ is spherically symmetric:
$$
  {\bf H'} = h(r){\bf r}.
$$
But this is impossible, as we noted above, because it is inconsistent
with our initial
Maxwell equations.

Thus our proof is finished.
\section{Conclusions}

To summarize, we have shown that the introduction of an arbitrary small
photon mass makes invalid the existing proofs of the consistency of the Dirac
monopole theory. More exactly, the massive electrodynamics does not allow any
generalization of the methods in which the Dirac monopole was introduced into
the massless electrodynamics. Not only the original Dirac scheme which
arrives at the quantization condition by using the gauge invariance,
single-valuedness of the wavefunction and "the veto" postulate does not work
anymore; but also the different approach relying on the algebra of angular
momentum fails in the case of massive electrodynamics. If the magnetic
monopole were ever to be introduced into massive electrodynamics
consistently, that would be possible only due to some radically new mechanism
compared with the existing ones.

What is the physical reason for that failure?

The whole existence of the Dirac monopole in the massless electrodynamics
rests upon the quantization condition which makes invisible the string
attached to the monopole. The quantization condition can be obtained either
with the help of gauge invariance or the angular momentum quantization. In
the massive case, both these approaches are not applicable anymore, as we
have shown. That means that there is hardly any way  to make the string
invisible in the massive electrodynamics.Thus, although the system ``string
plus monopole'' does exist in massive electrodynamics, it is very difficult,
if not impossible, to make a consistent theory of  the monopole {\em without}
a string.

One may think that our result contradict the principle of continuity which
states that any physical consequence of massive electrodynamics should go
smoothly into the corresponding result of the standard electrodynamics when
the photon mass tends to zero. Indeed, at first sight the appearance of the
Dirac monopoles at zero photon mass is an obvious discontinuity as compared
with their absence at an arbitrary small photon mass.

However, this simple argument is not yet sufficient to claim discontinuity.
An analogy with a similar ``discontinuity'' is instructive here: consider the
number of photon degrees of freedom in massive and massless electrodynamics.
the photon with a mass has three polarization states, independent of how
small its mass is. Then, as soon as the photon becomes massless, the
longitudinal polarization abruptly vanishes and we are left with only two
(transverse) polarization states. Does this fact create a discontinuity? No.
To find out if there is a discontinuity or not, we have  to study the
behavior of a more physical quantity, such as the probabilities to emit or
absorb a longitudinal photon, rather than merely counting the number of
degrees of freedom.

The analysis done by Schr\"{o}dinger shows that if one considers the
interaction of longitudinal photons with matter, this interaction vanishes as
photon mass tends to zero. Thus the longitudinal photons decouple in the
limit $ m_{\gamma} \rightarrow 0 $ so that the continuity is restored.

Coming back to our case with monopoles, one should carry out a similar
program to make sure the continuity is not violated.

Although we do not intend carrying out this program in the present work, we
would like to present here some physical arguments suggesting that the
continuity may be indeed preserved.  We have shown that instead of a true
Dirac monopole, massive electrodynamics contains a more cumbersome object
which can be viewed as consisting of three pieces:  the monopole, the string
and, finally, the additional "diffuse"magnetic field, eq.~ (\ref{bh}) (or the
corresponding vector potential, eq.~ (\ref{b}) ). Let us call this complex
object "a fake monopole". Before any discussion of the continuity problem,
one has of course make sure that fake monopoles can be described within a
logically consistent quantum theory. This is an open question beyond the
present paper, but for the sake of argument let us simply assume that the
answer is positive. Then, a natural question arises: how can we physically
distinguish between a fake and a true monopoles?

If the fake monopole has a "dequantized" magnetic charge (i.e., the charge
not obeying the Dirac quantization condition, eq.~ (\ref{dirac})) there does
not seem to be any problems. However, imagine that a fake monopole takes on a
magnetic charge which exactly coincides with one of the quantized values.
What happens then? Let us try and detect the string. One way to do that is to
use the Aharonov-Bohm effect. But then we run into trouble, because if the
magnetic charge has one of the Dirac values, {\em the strings \/} of a fake
and true monopoles are {\em exactly the same \/}. Therefore, {\em both of
them \/} are {\em invisible \/} in the Aharonov-Bohm type of experiment. The
way out of this difficulty is to remember about the existence of the feeble
"diffuse" component of the fake monopole which is of the order of $m^2$.
Generally speaking, this component would give non-zero contribution to the
Aharonov-Bohm effect and thus would, in principle, distinguish between a fake
and a true monopoles.

Thus, at least in a gedanken experiment we would be able to distinguish
between a fake and a true monopoles. (Whether this gedanken experiment can be
transformed into a real one, is an open question.) The observable difference
between the fake and true monopoles would vanish {\em smoothly \/} as the
photon mass tends to zero.

So we see that in this particular case there does not arise any problem of
discontinuity. It would be interesting to complement this type of analysis by
considering  other physical situations, such as Cabrera-type of experiment or
the scattering of the electron on a monopole.

On the other hand,  if the continuity is found to be broken (which we cannot
completely rule out at this stage), that would constitute a serious argument
against the existence of monopoles and thus would provide a possible
theoretical reason for the absence of monopoles in nature.

Alternatively, the continuous failure to discover the monopoles in the
experiment may be considered as an indirect evidence  for the finite photon
mass (unfortunately, we cannot say anything definite about the value of this
mass).

In this paper we have considered only fundamental monopoles within massive
U(1) electrodynamics. A natural question, then, is:  what about
t'Hooft-Polyakov monopoles which appear necessarily within gauge theories
based on simple groups (such as SU(5))? In such theories, one might give
small mass to the photon by spontaneous breaking of the electromagnetic U(1)
symmetry.

However, there is perhaps less interest in considering such kind of theories
because they seem to be ruled out by the experiment. More specifically, the
reason is that after giving photon a mass through spontaneous breaking of
U(1) symmetry  in such theories there arises very light charged particle
which is unacceptable experimentally \cite{we1,okun}. An important feature of
such theories is that the electric charge conservation is violated in these
theories: to give photon a mass, we need to have a {\em charged \/} scalar
field with {\em non-zero \/} vacuum expectation; this necessarily leads to
the electric charge non-conservation.

(To make our our discussion of this point more comprehensive, we note that
there does exist an acceptable way to spontaneously violate the electric
charge conservation and give photon a mass. But it requires introduction of
higgs particles with extremely small electric charge: for details, see
\cite{we1,we2}. However, even this mechanism is extremely difficult or
impossible to implement within theories based on semisimple groups (e.g.
SU(5)),  see the discussion in \cite{OVZ} ).

Thus the present work reveals  a new and rather general relation between the
two fundamental facts: the masslessness (massiveness?) of the photon and the
non-existence (existence?) of the magnetic monopole.

The authors are grateful to M.Shaposhnikov, R.Volkas and K.Wali for
stimulating discussions. We would like to thank H.Kleinert for drawing
our attention to Refs. \cite{k}.
This work was supported in part by the Australian Research Council.

\begin{figure}
\caption{The choice of coordinates for analysing the integral in Eq. (14).}
\end{figure}

\end{document}